\documentclass[aps,prd,twocolumn,nofootinbib,showpacs,superscriptaddress,floatfix]{revtex4-1}

\usepackage{amsmath}
\usepackage{epsfig}
\usepackage{graphicx}
\usepackage{rotating}

\newcommand\be{\begin{equation}}
\newcommand\ba{\begin{eqnarray}}
\newcommand\ee{\end{equation}}
\newcommand\ea{\end{eqnarray}}
\newcommand\nn{\nonumber}

\begin{document}

\title{Inner accretion disk edges in a Kerr-like spacetime}

\author{Tim Johannsen}
\affiliation{Department of Physics and Astronomy, University of Waterloo, Waterloo, Ontario N2L 3G1, Canada}
\affiliation{Canadian Institute for Theoretical Astrophysics, University of Toronto, Toronto, Ontario M5S 3H8, Canada}
\affiliation{Perimeter Institute for Theoretical Physics, Waterloo, Ontario N2L 2Y5, Canada}

\date{\today}

%%%%%%%%%%%%%%%%%%%%%%%%%%%%%%%%%%%%%%%%%%%%%%%%%
\begin{abstract}

According to the no-hair theorem, astrophysical black holes are uniquely described by the Kerr metric. In order to test this theorem with observations in either the electromagnetic or gravitational-wave spectra, several Kerr-like spacetimes have been constructed which describe potential deviations from the Kerr spacetime in parametric form. For electromagnetic tests of the no-hair theorem, such metrics allow for the proper modeling of the accretion flows around candidate black holes and the radiation emitted from them. In many of these models, the location of the inner edge of the accretion disk is of special importance. This inner edge is often taken to coincide with the innermost stable circular orbit, which can serve as a direct probe of the spin and the deviation from the Kerr metric. In certain cases, however, an innermost stable circular orbit does not exist, and the inner edge of an accretion disk is instead determined by an instability against small perturbations in the direction vertical to the disk. In this paper, I analyze the properties of accretion disks in the Kerr-like metric proposed by Johannsen and Psaltis [Phys. Rev. D {\bf 83}, 124015 (2011)], whose inner edges are located at the radii where this vertical instability occurs. I derive expressions of the energy and axial angular momentum of disk particles that move on circular equatorial orbits and calculate the locations of the inner disk edges. As a possible observable of such accretion disks, I simulate profiles of relativistically broadened iron lines and show that they depend significantly on the values of the spin and the deviation parameter.

\end{abstract}

\pacs{04.50.Kd,04.70.-s}

%04.30.Db Wave generation and sources 
% 04.50.Kd Modified theories of gravity
%04.70.-s	Physics of black holes 

\maketitle

%%%%%%%%%%%%%%%%%%%%%%%%%%%%%%%%%%%
%%%%%%%%%%%%%%%%%%%%%%%%%%%%%%%%%%%
%%%%%%%%%%%%%%%%%%%%%%%%%%%%%%%%%%%

\section{Introduction}
\label{sec:intro}

According to the no-hair theorem, black holes are uniquely characterized by their masses $M$ and spins $J$ and are described by the Kerr metric (e.g., \cite{Heusler96}). For this reason, all astrophysical black holes are expected to be Kerr black holes. Evidence for the validity of this assumption, however, has so far been inferred only indirectly (e.g., \cite{Narayan}). It is, therefore, desirable to test the no-hair theorem through observations.

Tests of the precise strong-field nature of black-hole candidates are often constructed in a model-independent, phenomenological approach, which does not assume any particular theory of gravity, and the underlying theory is generally unknown. Instead, these tests encompass large classes of modified theories of gravity, and insight into the fundamental theory is hoped to be gained through observations. It is then assumed that particles move along geodesics, which allows for the calculation and prediction of possible observable signatures of the theory. The assumption of geodesic particle motion is consistent with current experiments (see Ref.~\cite{WillLRR}) and holds in general relativity and in at least all modified theories of gravity with known black hole solutions that differ from the Kerr metric \cite{CSEDGB}. Several such strong-field tests have been suggested, which can be performed with observations in both the gravitational-wave~\cite{Ryan95,EMRI,kludge,CH04,GB06,Brink08,Gair08,Apostolatos09,VH10,Vigeland:2011ji,Gair:2011ym} and the electromagnetic spectra~\cite{PaperI,PaperII,PaperIII,EM,BambiDiskJet,BambiBarausse11,PJ11,PaperIV,BambiQPO,BambiIron}.

These strong-field tests rely on parametric spacetimes that deviate either slightly or severely from the Kerr metric, and several such Kerr-like metrics have been developed to date (e.g., \cite{MN92,CH04,GB06,VH10,JPmetric,Vigeland:2011ji}). See Ref.~\cite{Joh13} for a detailed discussion of the strong-field properties of these metrics. Since general relativity is still practically untested in the strong-field regime \cite{PsaltisLRR}, deviations from the Kerr metric, should they exist, could be either small or large as long as they are consistent with the limits imposed by current weak-field tests of general relativity (see Ref.~\cite{WillLRR}).

If a nonzero deviation from the Kerr metric is detected by either a gravitational-wave or an electromagnetic measurement, there are two possible interpretations. Within general relativity, the only possibility is that the compact object is not a black hole but instead either a stable stellar configuration or an exotic object~\cite{CH04,Hughes06}. Alternatively, black holes are different from the expected Kerr black holes (see, e.g., Refs.~\cite{YP09,SY09,Psaltis08}), and general relativity is modified in the strong-field regime.

Other tests of the no-hair theorem exist in both the gravitational-wave and electromagnetic spectra. After a merger with an object of comparable mass, such a perturbed black hole in general relativity quickly settles back down to the stationary Kerr solution and, in the process, emits characteristic gravitational ringdown radiation in the form of so-called quasinormal modes. These modes are uniquely determined by the mass and spin of the black hole and detection of at least two of these modes, therefore, tests whether the end state of the merger is a Kerr black hole constituting another strong-field test of general relativity \cite{ringdown}. Electromagnetic tests also include the monitoring of close stellar orbits around Sgr~A* \cite{Sgr} and of pulsar black-hole binaries \cite{pulsars}, which are both weak-field tests.

Strong-field tests of the no-hair theorem in the electromagnetic spectrum are based on the imprints of the underlying spacetime on the emission from the accretion flows around black-hole candidates, which can alter the observed signals in a characteristic manner \cite{PaperI}. Possible observables include the direct imaging of supermassive black holes with very-long baseline interferometric techniques \cite{PaperII}, relativistically broadened iron lines \cite{PJ11,PaperIV,BambiIron}, quasiperiodic variability \cite{PaperIII,PaperIV,BambiQPO}, thermal accretion disk spectra \cite{BambiBarausse11,BambiDiskJet,Krawc12}, x-ray polarization \cite{Krawc12}, and, potentially, jets \cite{BambiDiskJet}.

In many of these tests, accretion flows are modeled as thin disks that lie in the equatorial plane of the compact object and that can be either optically thin or optically thick. In such models, the particles in the disk typically move on circular orbits. Deviations from the Kerr metric can have a strong effect primarily on the flux emitted in the part of such disks that is closest to the compact object, because the spacetime curvature experienced by the disk is largest at the inner edge. The inner edge of a disk around a black hole candidate is, many times, taken to coincide with the innermost stable circular orbit (ISCO). Since the temperature of a thermal disk spectrum reaches its maximum near the ISCO (e.g., \cite{Zhang97}), this inner region of an accretion disk can contribute a significant portion to the total observed disk flux.

While the location of the ISCO in the Kerr metric (measured in units of the mass $M$) depends only on the value of the spin, the location of the ISCO usually depends on both the spin and the deviation parameters in a Kerr-like metric (e.g., \cite{PaperI}). At the ISCO radius, the radial epicyclic frequency vanishes, and circular equatorial orbits are radially unstable inside of the ISCO (the so-called plunging region). In certain Kerr-like metrics such as the quasi-Kerr \cite{GB06} and Manko-Novikov \cite{MN92} metrics, however, for certain values of the deviation parameters, all circular equatorial orbits are radially stable. Instead, the vertical epicyclic frequency vanishes at a radius $r\sim {\rm a~few}~M$, and such orbits experience an instability in the vertical direction \cite{Gair08,PaperIII}. Very close to the origin, this vertical instability may disappear, and circular equatorial orbits are, again, stable \cite{Gair08}, which may lead to characteristic gravitational-wave emission during an extreme mass-ratio inspiral \cite{Apostolatos09,BLG13}. As was shown in Ref.~\cite{BambiBarausse2}, the vertical instability of circular equatorial orbits in the Manko-Novikov metric can cause geometrically thin accretion disks to become geometrically thick in the innermost region and the disk plasma may get trapped inside of the inner disk edge. The orbital stability of these particles, in turn, depends sensitively on their energies and angular momenta, which may lead to an outflow of disk plasma in the direction vertical to the disk \cite{CHL12,LGC13}.

In this paper, I study the properties of circular equatorial orbits and the inner edges of thin accretion disks in the Kerr-like metric constructed by Johannsen and Psaltis \cite{JPmetric}. This metric depends on a set of free parameters and derives from a general modification of the $(t,t)$ and $(r,r)$ components of the Schwarzschild metric with a successive transformation of the Newman-Janis type \cite{NewmanJanis}. In general relativity, the Newman-Janis transformation is a complex rotation that transforms the Schwarzschild metric into the Kerr metric \cite{NewmanJanis}. While this transformation generally does not relate static solutions to the corresponding stationary solutions in modified gravity theories, it can be used to introduce spin to modified Schwarzschild metrics (see, also, Refs.~\cite{VH10,Vigeland:2011ji}). If all deviation parameters vanish, such a transformed static metric, then, reduces smoothly to the Kerr metric. Contrary to other Kerr-like metrics such as the quasi-Kerr \cite{GB06} and Manko-Novikov \cite{MN92} metrics, which often contain singularities or closed timelike curves outside of the central object that can limit the utility of a given metric to model accretion flows around black-hole candidates, the metric by Johannsen and Psaltis is free of such pathologies \cite{JPmetric,Joh13}. 

In this metric, an ISCO exists for most deviations from the Kerr metric, while circular equatorial orbits become vertically unstable if an ISCO does not exist \cite{JPmetric}. In the former case, the location of the ISCO, as well as certain properties of circular equatorial orbits, was already analyzed in Ref.~\cite{JPmetric}. Here I focus on the case when an ISCO does not exist. First I derive expressions of the energy and axial angular momentum of a particle on a circular equatorial orbit, as well as the radial and vertical epicyclic frequencies. I calculate the radial locations, at which circular equatorial orbits become unstable in the vertical direction, and show that these locations can be shifted significantly as a function of the spin and the deviation from the Kerr metric. I compare these radial locations to those of the ISCO in the part of the parameter space where an ISCO exists.

As a possible observable of accretion disks, whose inner edges are determined by the vertical instability of circular equatorial orbits instead of the location of the ISCO, I simulate profiles of relativistically broadened iron lines (see Refs.~\cite{PaperIV,BambiIron} for a discussion of iron lines in this metric emitted from disks that terminate at the ISCO). I show that these profiles can depend significantly on the deviation from the Kerr metric, which affects primarily the flux ratio of the two peaks in the double-horn shape of the line profile, as well as the low-energy part of the line. While at a given value of the spin the ISCO radius decreases for increasing values of the deviation parameter, the radius at which the vertical instability occurs likewise increases. This effect results in a different dependence of the profile shape on the deviation parameter and, in particular, in different flux ratios of the two peaks in iron lines with unequal values of the deviation parameter for which the ISCO and the radius of the vertical instability are located at the same radius. I also show, however, that profiles for different choices of the spin and the deviation parameter that correspond to the same location of the inner edge as determined by the vertical instability are very similar.

\begin{figure*}[ht]
\begin{center}
\psfig{figure=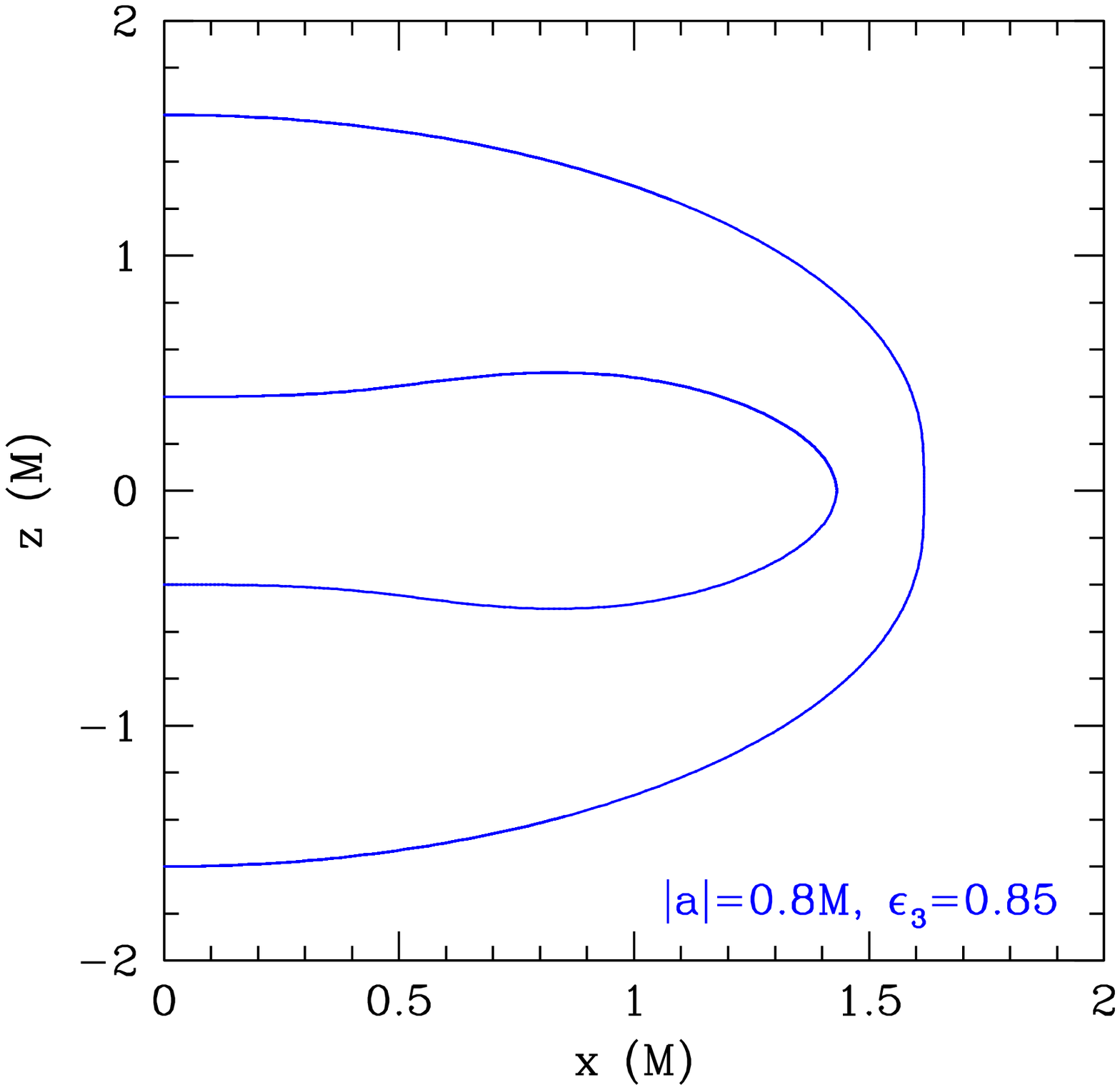,width=0.32\textwidth}
\psfig{figure=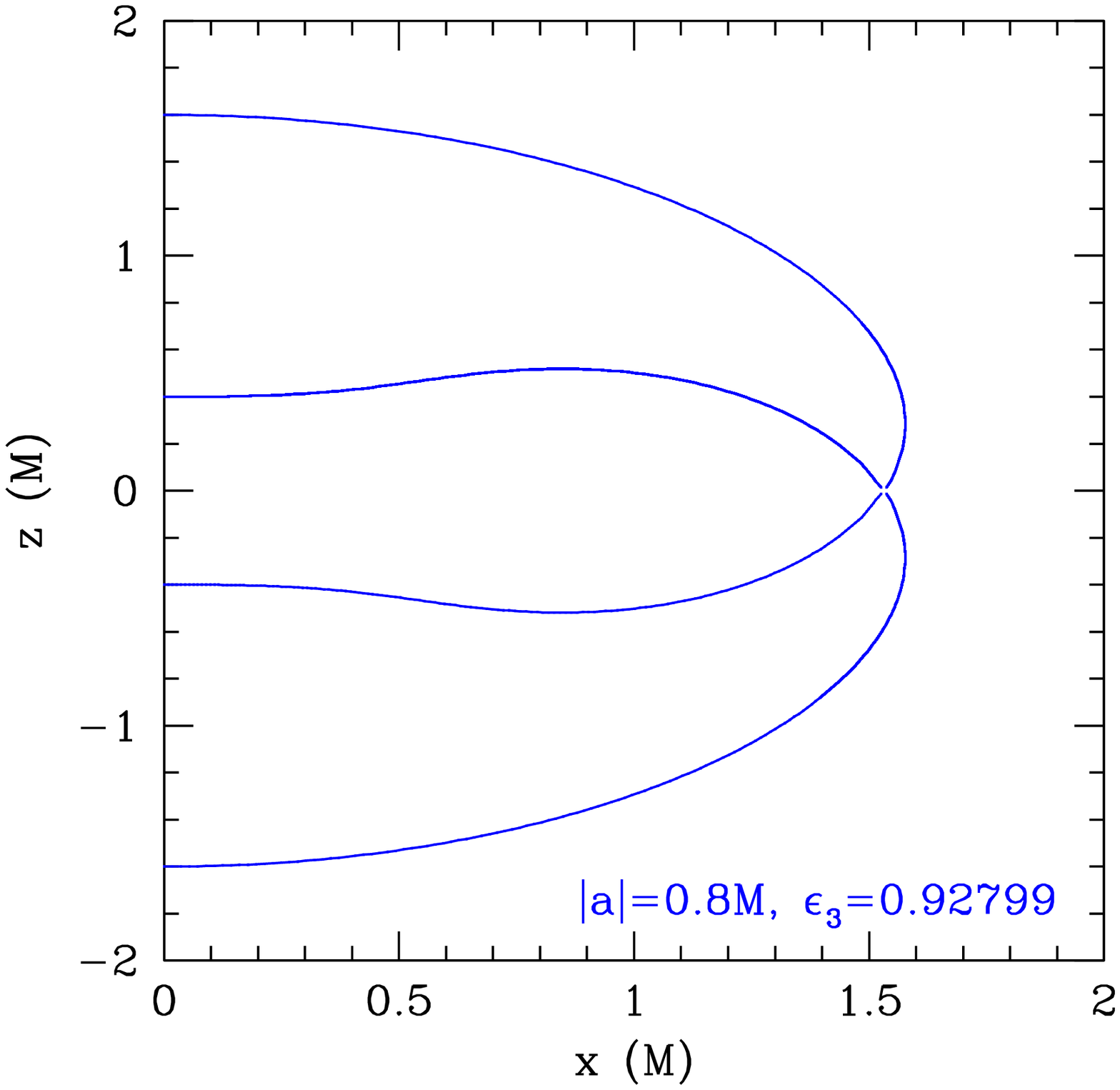,width=0.32\textwidth}
\psfig{figure=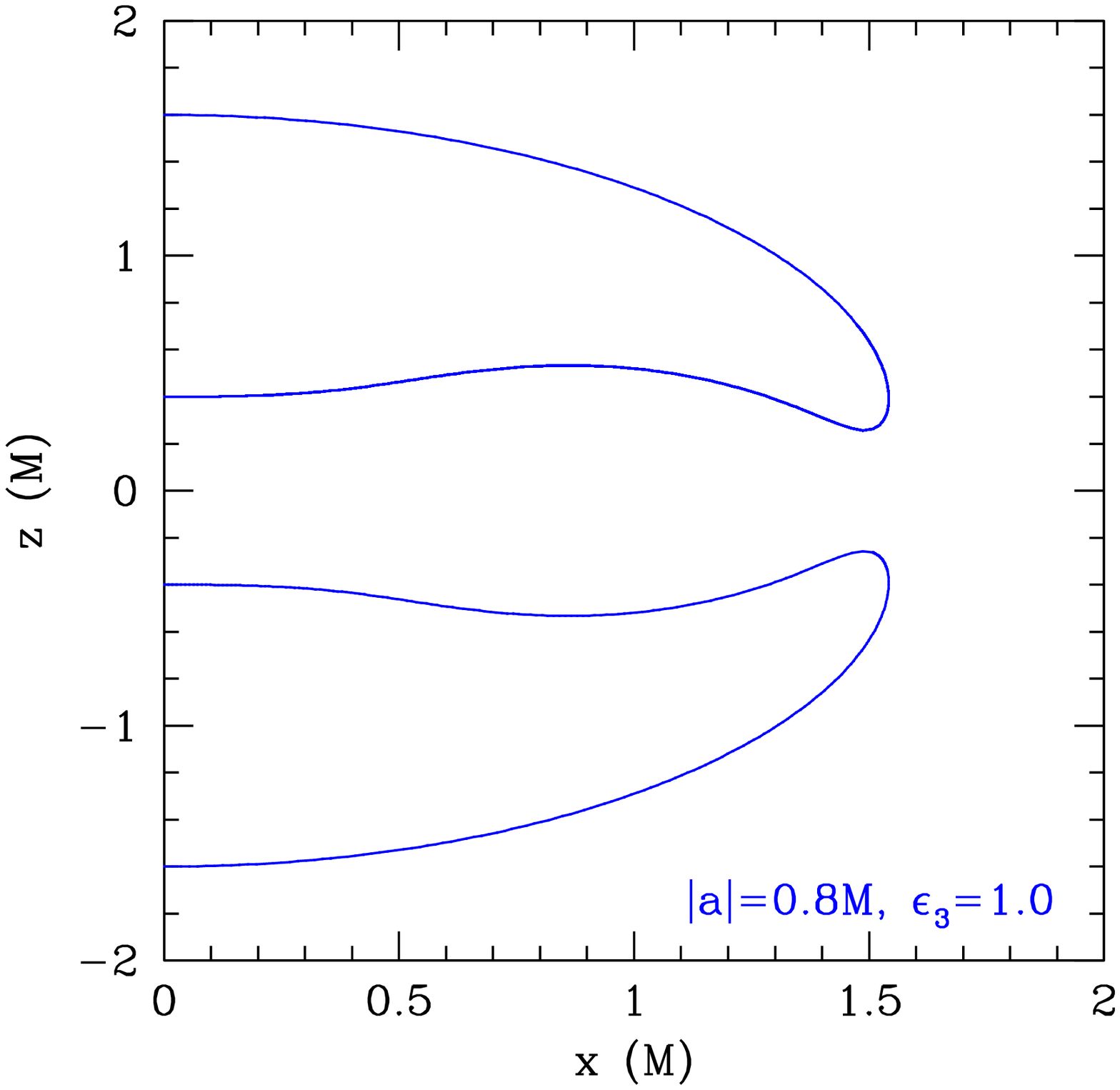,width=0.32\textwidth}
\end{center}
\caption{Different shapes of the (singular) Killing horizon for values of the spin $|a|=0.8M$ and deviation parameter $\epsilon_3$. Left: $\epsilon_3=0.85<\epsilon_3^{\rm bound}(0.8M)$, where $\epsilon_3^{\rm bound}(a)$ denotes the boundary between the regions of the parameter space with different topologies of the Killing horizon. The inner and outer Killing horizons have spherical topology. Center: $\epsilon_3=0.92799\approx\epsilon_3^{\rm bound}(0.8M)$. The inner and outer Killing horizons merge in the equatorial plane. Right: $\epsilon_3=1.0>\epsilon_3^{\rm bound}(0.8M)$. The inner and outer Killing horizons have split into two spherelike surfaces located above and below the equatorial plane, respectively. For all values of the parameter $\epsilon_3$, the origin is likewise singular.}
\label{fig:topo}
\end{figure*}

In Sec.~\ref{sec:metric}, I discuss some of the properties of the metric \cite{JPmetric} and derive the energy and axial angular momentum of a particle on a circular equatorial orbit. I analyze the stability of inner accretion disk edges in this metric in Sec.~\ref{sec:edges}. In Sec.~\ref{sec:lines}, I simulate iron line profiles, and I discuss the results in Sec.~\ref{sec:discussion}. Throughout this paper, I use geometric units and set $G=c=1$, where $G$ and $c$ are the gravitational constant and the speed of light, respectively.

%%%%%%%%%%%%%%%%%%%%%%%%%%%%%%%%%%%
\section{Metric Properties}
\label{sec:metric}

In this section, I discuss some of the properties of the metric constructed by Johannsen and Psaltis \cite{JPmetric} and calculate the energy and axial angular momentum of a particle on a circular equatorial orbit.

In Boyer-Lindquist-like coordinates, i.e., in spherical-like coordinates $(t,r,\theta,\phi)$ which reduce to the standard Boyer-Lindquist coordinates if all deviations from the Kerr metric vanish, the metric is given by the expressions
\ba
g_{tt} &=& -[1+h(r,\theta)] \left(1-\frac{2Mr}{\Sigma}\right), \nn \\
g_{t\phi} &=& -\frac{ 2aMr\sin^2\theta }{ \Sigma }[1+h(r,\theta)], \nn \\
g_{rr} &=& \frac{ \Sigma[1+h(r,\theta)] }{ \Delta + a^2\sin^2\theta h(r,\theta) }, \nn \\
g_{\theta\theta} &=& \Sigma, \nn \\
g_{\phi\phi} &=& \left[ \sin^2\theta \left( r^2 + a^2 + \frac{ 2a^2 Mr\sin^2\theta }{\Sigma} \right) \right. \nn \\
&& ~ + \left. h(r,\theta) \frac{a^2(\Sigma + 2Mr)\sin^4\theta }{\Sigma} \right],
\label{eq:metric}
\ea
where
\ba
\Delta &\equiv& r^2-2Mr+a^2, \\
\Sigma &\equiv& r^2+a^2\cos^2 \theta,
\label{eq:deltasigma}
\ea
and where
\be
h(r,\theta) \equiv \sum_{k=1}^\infty \left( \epsilon_{2k} + \epsilon_{2k+1}\frac{Mr}{\Sigma} \right) \left( \frac{M^2}{\Sigma} \right)^{k}
\label{eq:hfull}
\ee
is a function that contains the deviation parameters $\epsilon_k$. In this paper, I will use this metric with only one nonzero parameter, so that the function $h(r,\theta)$ reduces to
\be
h(r,\theta) = \epsilon_3 \frac{M^3 r}{\Sigma^2}.
\label{eq:h}
\ee
In the limit $\epsilon_3\rightarrow0$, this metric reduces to the Kerr metric.

As shown in Ref.~\cite{Joh13}, if $\epsilon_3\neq0$, this metric harbors a naked singularity, which is located at the Killing horizon. For small deviations from the Kerr metric, if the metric elements given by Eq.~\eqref{eq:metric} are expanded to first order in the parameter $\epsilon_3$, this singularity disappears, and the metric describes a black hole.

While the Killing horizon has spherical topology for most values of the spin and the deviation parameter, for values of the parameter $\epsilon_3\geq\epsilon_3^{\rm bound}$, where
\ba
\epsilon_3^{\rm bound}(a) \equiv && \frac{1}{3125(a/M)^2} \bigg[ 1024 \left( 4 + \sqrt{16-15(a/M)^2} \right) \nn \\
&& - 160(a/M)^2 \left(40+7\sqrt{16-15(a/M)^2} \right) \nn \\
&& + 150 (a/M)^4 \left(15+\sqrt{16-15(a/M)^2} \right) \bigg],
\label{eq:ep3bound}
\ea
the topology of the Killing horizon is disjoint consisting of two spherical surfaces located above and below the equatorial plane and centered at the symmetry axis (except for $a=\pm M$, which corresponds to the extremal Kerr black holes) \cite{Joh13}. In Fig.~\ref{fig:topo}, I plot the location of the Killing horizon in the $xz$ plane, where $x \equiv \sqrt{r^2+a^2}\sin\theta$ and $z \equiv r\cos\theta$, for a value of the spin $a=0.8M$ and several values of the parameter $\epsilon_3$. As shown in Ref.~\cite{JPmetric}, the topology of the Killing horizon is also related to the existence of an ISCO in this spacetime. An ISCO exists for values of the spin $a\leq0$ as well as for values of the spin $a>0$ and values of the parameter $\epsilon_3<\epsilon_3^{\rm bound}$ (as long as $\epsilon_3\lesssim30$). 

The properties of a particle on a circular equatorial orbit depend directly on its orbital energy and angular momentum. For the case when $\epsilon_3<\epsilon_3^{\rm bound}$, these expressions were calculated in Ref.~\cite{JPmetric} and were written in terms of several functions $P_i$, $i=1,...,6$, which are polynomials in the radial coordinate $r$ apart from a few square root terms. As was pointed out in Ref.~\cite{Krawc12}, a sign change can occur in the axial angular momentum $L_z$ at a radius very close to the central object, where the function $P_5$ has a root. Circular equatorial orbits likewise exist in the part of the parameter space, where $\epsilon_3\geq\epsilon_3^{\rm bound}$. In the following, I generalize the expressions of the energy and axial angular momentum of a particle on a circular equatorial orbit that were derived in Ref.~\cite{JPmetric} to include the entire parameter space. The resulting expressions of the energy and axial angular momentum of particles on such orbits take the full root structure of the functions $P_1$ to $P_6$ into account.

The derivation is very similar to the one in Ref.~\cite{JPmetric}. Denoting the 4-momentum of the particle by
\be
p^\alpha = \mu \frac{dx^\alpha}{d\tau},
\ee
where $\mu$ is its rest mass, the energy $E=-p_t$ and axial angular momentum $L_z=p_\phi$ are conserved along its orbit. Solving the equation
\be
p^\alpha p_\alpha = -\mu^2
\ee
in the equatorial plane for the radial and vertical momenta and inserting the constants of the motion, I obtain the effective potential
\ba
&& V_{\rm eff}(r) \equiv \frac{1}{2}\mu^2\left[ g_{rr}\left(\frac{dr}{d\tau}\right)^2 + g_{\theta\theta} \left(\frac{d\theta}{d\tau}\right)^2 \right]\nn \\
&& = - \frac{1}{2} (g^{tt}E^2 -2g^{t\phi}EL_z + g^{\phi\phi}L_z^2 + \mu^2),
\ea
which governs the motion of particles in the equatorial plane.

Particle motion is only allowed if $V_{\rm eff}(r)\geq0$, which I can express in terms of the constants of motion. At a given radius $r$ and for a given axial angular momentum $L_z$, particle motion is only allowed for values of the energy $E\geq E_{\rm min}$. From the equation
\be
V_{\rm eff}(r) = 0, \label{Veff}
\ee
I obtain the minimum energy
\begin{widetext}
\ba
E_{\rm min} &=& \frac{1}{r^2 [r^6+a^2(r+2M)(r^3+\epsilon_3M^3)]} \bigg[ 2aMr^2L_z(r^3+\epsilon_3M^3) \nn \\
&& + \sqrt{ r^2(r^3+\epsilon_3M^3)[r^4(r-2M)+a^2(r^3+\epsilon_3M^3)]\{r^4L_z^2+[r^6+a^2(r+2M)(r^3+\epsilon_3M^3)]\mu^2\} } \bigg].
\ea
\end{widetext}
Smaller values of the energy are excluded, because these correspond to negative values of the energy as observed in the inertial frame of a local observer. Note, however, that the minimum energy (as observed at radial infinity) can be negative.

Solving the equation
\be
\frac{dV_{\rm eff}(r)}{dr} = 0 \label{dVeff} 
\ee
for the axial angular momentum and combining the result with Eq.~\eqref{Veff}, I derive the expressions
\ba
\frac{E}{\mu} &=& \sigma_1 \frac{1}{r^6}\sqrt{ \frac{P_1 + P_2}{P_3} },
\label{eq:E} \\
\frac{L_z}{\mu} &=& \frac{1}{r^4 P_6} \bigg[ \sigma_2 \sqrt{\frac{P_5}{P_3} } \nn \\
&& + \sigma_3 6a M (r^3 + \epsilon_3 M^3) \sqrt{ \frac{P_1 + P_2}{P_3} } \bigg],
\label{eq:Lz}
\ea
where $\sigma_{1-3}\equiv\pm1$. These expressions are valid for all values of the spin $|a|\leq M$ and the parameter $\epsilon_3$, provided that a circular equatorial orbit exists.

\begin{figure}[ht]
\begin{center}
\psfig{figure=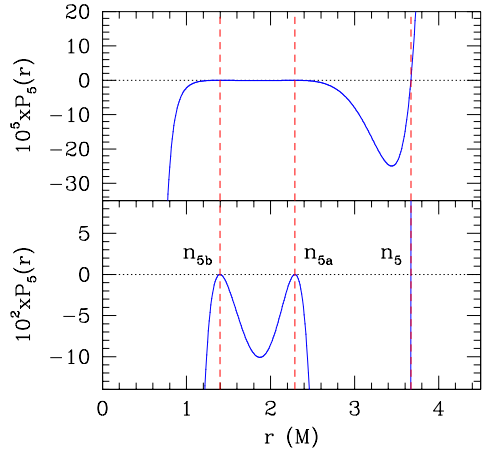,width=0.4\textwidth}
\end{center}
\caption{$P_5(r)$ as a function of the radius $r$ with parameters $a=0.7M$ and $\epsilon_3=2.0$. Top panel: The function $P_5(r)$ is very steep and has several local minima and maxima as well as five roots marked by red vertical dashed lines, one single root labeled ``${\rm n_5}$'' and two double roots labeled ``${\rm n_{5a}}$'' and ``${\rm n_{5b}}$,'' respectively. Lower panel: Enlarged view of the function $P_5$ around the roots on the $r$ axis.}
\label{fig:P5roots}
\end{figure}

\begin{figure}[ht]
\begin{center}
\psfig{figure=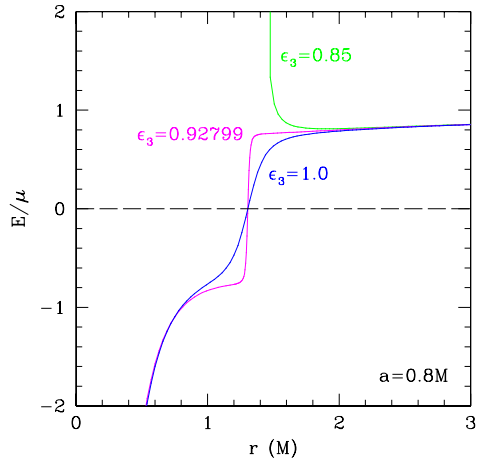,width=0.4\textwidth}
\psfig{figure=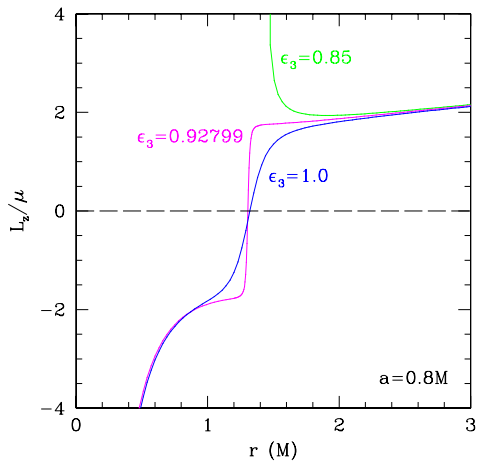,width=0.4\textwidth}
\end{center}
\caption{(Energy $E$ (top panel) and axial angular momentum $L_z$ (bottom panel) in units of the rest mass of a particle on a circular equatorial orbit as observed at radial infinity for a spin value $a=0.8M$ and several values of the parameter $\epsilon_3$. If the Killing horizon has spherical topology [i.e., if $\epsilon_3<\epsilon_3^{\rm bound}(0.8M)\approx0.92799$], both $E$ and $L_z$ diverge to plus infinity at the horizon and are imaginary inside. Otherwise, they change sign near the outermost projected location of the Killing horizon onto the equatorial plane and diverge to minus infinity at the origin.}
\label{fig:ELz}
\end{figure}

The signs of the constants $\sigma_{1-3}$ depend on the (real) root structure of the functions $P_1$ to $P_5$. In this paper, I only discuss the real roots of these functions if they lie in the range $r>0$ as well as outside of the ISCO, if an ISCO exists. Only in this case are these roots relevant for the discussion of circular equatorial orbits. It turns out that the roots of the polynomial $P_6$, which occurs in the denominator of the axial angular momentum, play no role. If such a root exists in the range that is relevant for circular equatorial orbits, the numerator of Eq.~\eqref{eq:Lz} likewise vanishes, and the expression of the axial angular momentum is smooth across the root. As an example of the roots of the functions $P_1$ to $P_5$, I plot the function $P_5(r)$ in Fig.~\ref{fig:P5roots} for values of the spin $a=0.7M$ and the parameter $\epsilon_3=2.0$. The function depends very sensitively on the radius $r$ and is very steep at most radii. In addition to several local maxima and minima, the function contains five roots, four of which are double roots.

For most circular orbits, the expressions of the energy and axial angular momentum given by Eqs.~\eqref{eq:E} and \eqref{eq:Lz} are identical to the ones in Ref.~\cite{JPmetric} (apart from a minor rearrangement of the function $P_3$), i.e., $\sigma_1=+1$, $\sigma_2=\pm1$, and $\sigma_3=\mp1$, where the upper signs refer to prograde orbits, while the lower signs refer to retrograde orbits. For values of the parameter $-10<\epsilon_3<10$, sign changes can only occur at radii $r\lesssim4.5M$, most of which lie inside of the ISCO if an ISCO exists. These sign changes usually only affect the energy and axial angular momentum slightly if an ISCO exists.

If $\epsilon_3<\epsilon_3^{\rm bound}$, $\sigma_1=+1$. If $\epsilon_3 \geq \epsilon_3^{\rm bound}$, the value of $\sigma_1$ depends on the existence of a root of the function $P_1+P_2$. The values of $\sigma_2$ and $\sigma_3$ depend on the root structure of all functions $P_1$ to $P_5$, and several different cases have to be considered. Outside of the Killing horizon, if the function $P_3$ has roots, the functions $P_1+P_2$ and $P_5$ have roots that coincide with the roots of the function $P_3$, and the expressions of the energy and axial angular momentum remain finite at the roots. If the Killing horizon has spherical topology, the expressions of the energy and axial angular momentum become infinitely large at the singularity. If the Killing horizon is disjoint and there is no naked singularity in the equatorial plane apart from the origin, the energy and axial angular momentum change sign at a radius $r\sim M$, and $E,L_z \rightarrow -\infty$ as $r\rightarrow0$. In Fig.~\ref{fig:ELz}, I plot the energy and axial angular momentum for a value of the spin $a=0.8M$ for several values of the parameter $\epsilon_3$, where the Killing horizon has spherical topology for values of the parameter $\epsilon_3<\epsilon_3^{\rm bound}\approx0.92799$ and disjoint topology for values of the parameter $\epsilon_3\geq \epsilon_3^{\rm bound}$. I list the full expressions of the energy and axial angular momentum given the roots of the functions $P_1$ to $P_5$ in the Appendix.

Note that in certain cases both the energy and axial angular momentum as observed at radial infinity can be negative. While negative values of the axial angular momentum simply correspond to retrograde orbits, negative values of the energy are best interpreted in the inertial frame of a local observer (see discussion in Ref.~\cite{BPT72}). For such a frame I can choose an orthonormal basis with vectors
\be
e_{\hat\alpha} = e_{~\hat\alpha}^\mu e_\mu, 
\ee
which obey the relations
\be
e_{~\hat\alpha}^\mu e_{~\hat\beta}^\nu g_{\mu\nu} = \eta_{{\hat\alpha}{\hat\beta}},
\label{eq:ONB}
\ee
where $\eta_{{\hat\alpha}{\hat\beta}}={\rm diag}(-1,1,1,1)$ is the Minkowski metric. Making the ansatz for the basis vectors
\ba
e_{\hat t} &=& \eta e_t + \gamma e_\phi, \nn \\
e_{\hat r} &=& \frac{1}{\sqrt{g_{rr}}} e_r, \nn \\
e_{\hat \theta} &=& \frac{1}{\sqrt{g_{\theta\theta}}} e_\theta, \nn \\
e_{\hat\phi} &=& \frac{1}{\sqrt{g_{\phi\phi}}} e_\phi,
\label{eq:localframe}
\ea
I use the relations given by Eq.~\eqref{eq:ONB} to solve for the parameters $\eta$ and $\gamma$. I obtain the solutions
\ba
\eta &=& \sqrt{ \frac{ g_{\phi\phi} }{ g_{t\phi}^2 - g_{tt}g_{\phi\phi} } } \nn \\
&=& \sqrt{ \frac{ r^2[r^6+a^2(r+2M)(r^3+\epsilon_3M^3)] }{ (r^3+\epsilon_3M^3)[r^4(r-2M)+a^2(r^3+\epsilon_3M^3)] } }, \nn \\
\gamma &=& -\frac{g_{t\phi}}{g_{\phi\phi}} \sqrt{ \frac{ g_{\phi\phi} }{ g_{t\phi}^2 - g_{tt}g_{\phi\phi} } } \nn \\
&=& -2aMr \bigg\{ \frac{ r^3+\epsilon_3M^3 }{ [r^6+a^2(r+2M)(r^3+\epsilon_3M^3)] } \nn \\
&& \times \frac{1}{ [r^4(r-2M)+a^2(r^3+\epsilon_3M^3)] } \bigg\}^{\frac{1}{2}}.
\label{eq:deltagamma}
\ea

The locally measured energy and axial angular momentum are the projections of the 4-momentum of the particle onto the basis vectors $e_{\hat t}$ and $e_{\hat\phi}$ given by Eq.~\eqref{eq:localframe}:
\ba
p^{\hat t} &=& \eta E - \gamma L_z, \\
p^{\hat \phi} &=& \frac{1}{\sqrt{g_{\phi\phi}}} L_z.
\ea

Solving these equations, I can write the energy as
\be
E = \frac{1}{\eta} \left[ p^{\hat t} + \sqrt{g_{\phi\phi}} \gamma p^{\hat \phi} \right].
\label{eq:Elocal}
\ee
Since all physical particles must have momenta with
\be
p^{\hat t} > \left| p^{\hat \phi} \right|,
\ee
negative values of the energy can only occur if 
\be
1 - \frac{g_{t\phi}^2}{g_{t\phi}^2-g_{tt}g_{\phi\phi}} > 0,
\ee
i.e.,
\be
g_{tt} > 0,
\ee
where I used Eqs.~\eqref{eq:deltagamma}. Therefore, negative values of the energy as observed at radial infinity can only occur in the ergosphere.

\begin{figure}[ht]
\begin{center}
\psfig{figure=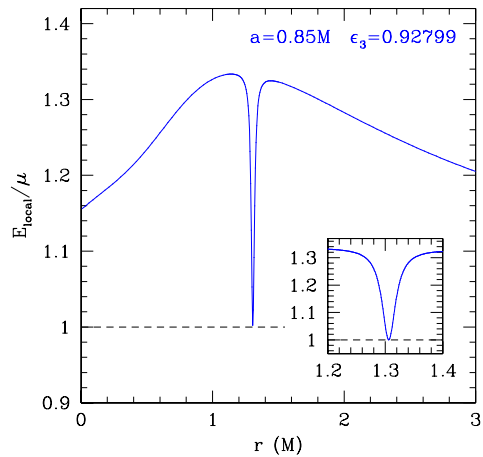,width=0.4\textwidth}
\end{center}
\caption{Energy $E_{\rm local}$ in units of the rest mass of a particle on a circular equatorial orbit as observed by a local observer for values of the spin $a=0.8M$ and the parameter $\epsilon_3=0.92799$. The energy is positive at all radii $r>0$. In particular, at least at the radii shown in this figure, the energy is greater than or equal to the rest mass of the particle, where equality holds only at a radius $r\approx1.31M$ (see inset).}
\label{fig:Elocal}
\end{figure}

In Fig.~\ref{fig:Elocal}, I plot the energy as a function of the radius for values of the spin $a=0.8M$ and the parameter $\epsilon_3=0.92799$ as observed in the local frame given by Eqs.~\eqref{eq:localframe}. While the energy observed at radial infinity is negative at small radii (c.f., Fig.~\ref{fig:ELz}), it is always positive in the local frame.

%%%%%%%%%%%%%%%%%%%%%%%%%%%%%%%%%%%%%%%%%%%%%%%%%%%%%

\section{Inner Accretion Disk Edges}
\label{sec:edges}

For values of the spin $a\leq0$ as well as of the spin $a>0$ and the parameter $\epsilon_3<\epsilon_3^{\rm bound}$, circular equatorial orbits are radially unstable inside of the ISCO radius, while circular equatorial orbits are radially stable for values of the spin $a>0$ and the parameter $\epsilon_3\geq\epsilon_3^{\rm bound}$. In the latter region, circular equatorial orbits are, instead, unstable against small perturbations in the direction vertical to the equatorial plane \cite{JPmetric}. At this radius, accretion disks are likely to acquire a certain thickness and may be disrupted (see Refs.~\cite{BambiBarausse2,CHL12,LGC13}). Near the origin, the vertical instability disappears, and a second region of stable circular orbits emerges inside of this radius, which extends down to the origin. In principle, this region could harbor a small inner part of the accretion disk, which is disconnected from the accretion disk located at radii outside of the vertical instability (see Refs.~\cite{Gair08,Q11,BLG13}).

In order to determine the radial locations where circular equatorial motion is either stable or unstable in the radial or vertical directions, I calculate the radial and vertical epicyclic frequencies $\Omega_r$ and $\Omega_\theta$ following the derivation in Ref.~\cite{PaperIII}. From the effective potential given by Eq.~\eqref{Veff}, radial and vertical motions around a circular equatorial orbit at a radius $r_0$ are governed by the equations
\ba
\frac{1}{2} \left( \frac{dr}{dt} \right)^2 &=& \frac{ V_{\rm eff} }{ g_{\rm rr}(p^{\rm t})^2 } \equiv V_{\rm eff}^r, \label{Veffr} \\
\frac{1}{2} \left( \frac{d\theta}{dt} \right)^2 &=& \frac{ V_{\rm eff} }{ g_{\rm \theta\theta}(p^{\rm t})^2 } \equiv V_{\rm eff}^{\rm \theta},
\label{Vefftheta}
\ea
respectively.

I then introduce small perturbations $\delta r$ and $\delta \theta$ and take the coordinate time derivative of Eqs. (\ref{Veffr}) and (\ref{Vefftheta}), which yields
\ba
\frac{d^2(\delta r)}{dt^2} &=& \frac{ d^2 V_{\rm eff}^r }{ dr^2 } \delta r, \\
\frac{d^2(\delta \theta)}{dt^2} &=& \frac{ d^2 V_{\rm eff}^{\rm \theta} }{ d\theta^2 } \delta \theta.
\ea
From these expressions, I derive the radial and vertical epicyclic frequencies as
\ba
\Omega_r^2 &=& -\frac{ d^2 V_{\rm eff}^r }{dr^2}, \\
\label{kappa}
\Omega_{\rm \theta}^2 &=& -\frac{ d^2 V_{\rm eff}^{\rm \theta} }{ d\theta^2 },
\label{omegatheta}
\ea
where the second derivatives are evaluated at $r=r_0$. These expressions are lengthy, and I do not write them here explicitly. 

\begin{figure}[ht]
\begin{center}
\psfig{figure=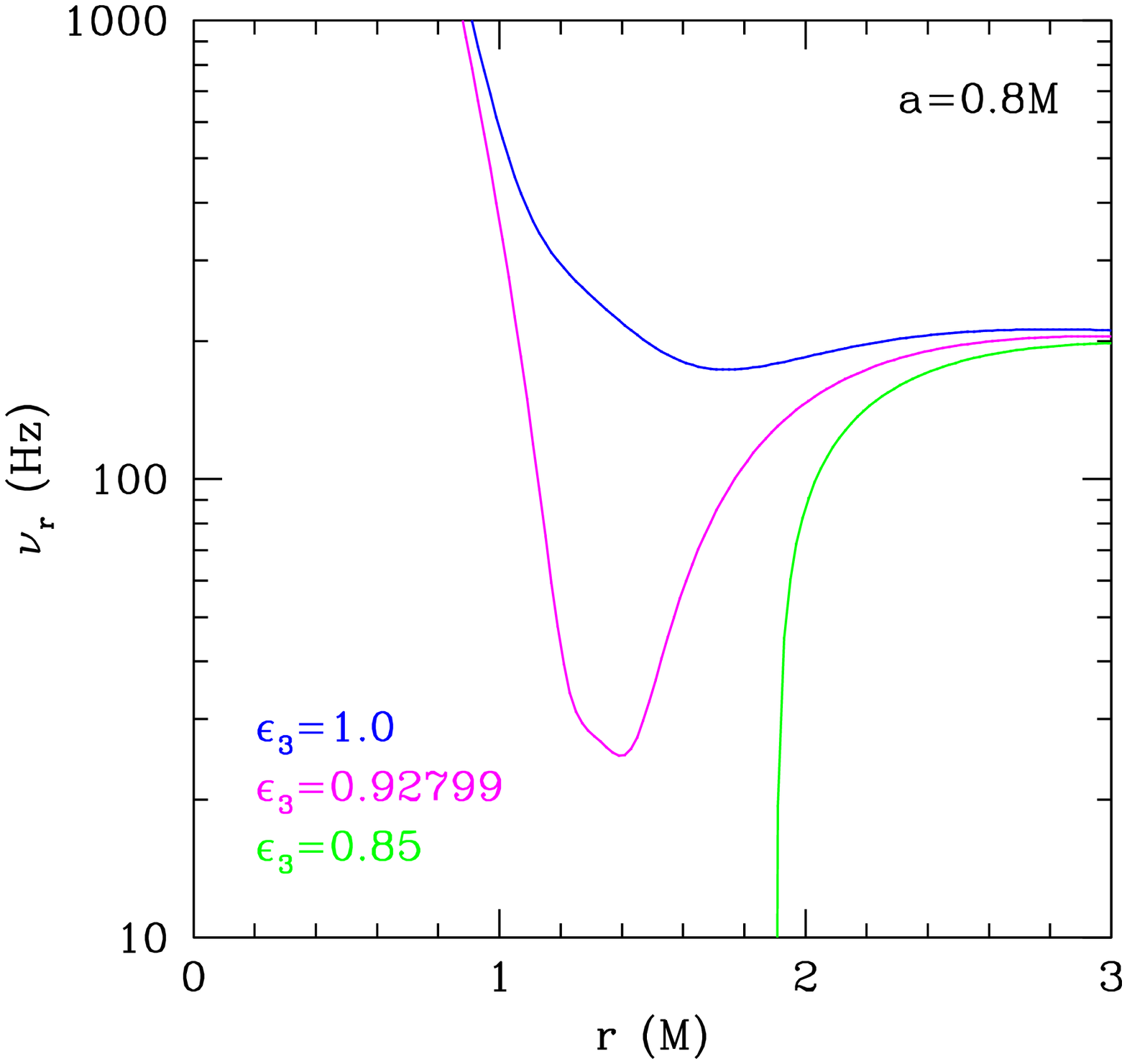,width=0.4\textwidth}
\psfig{figure=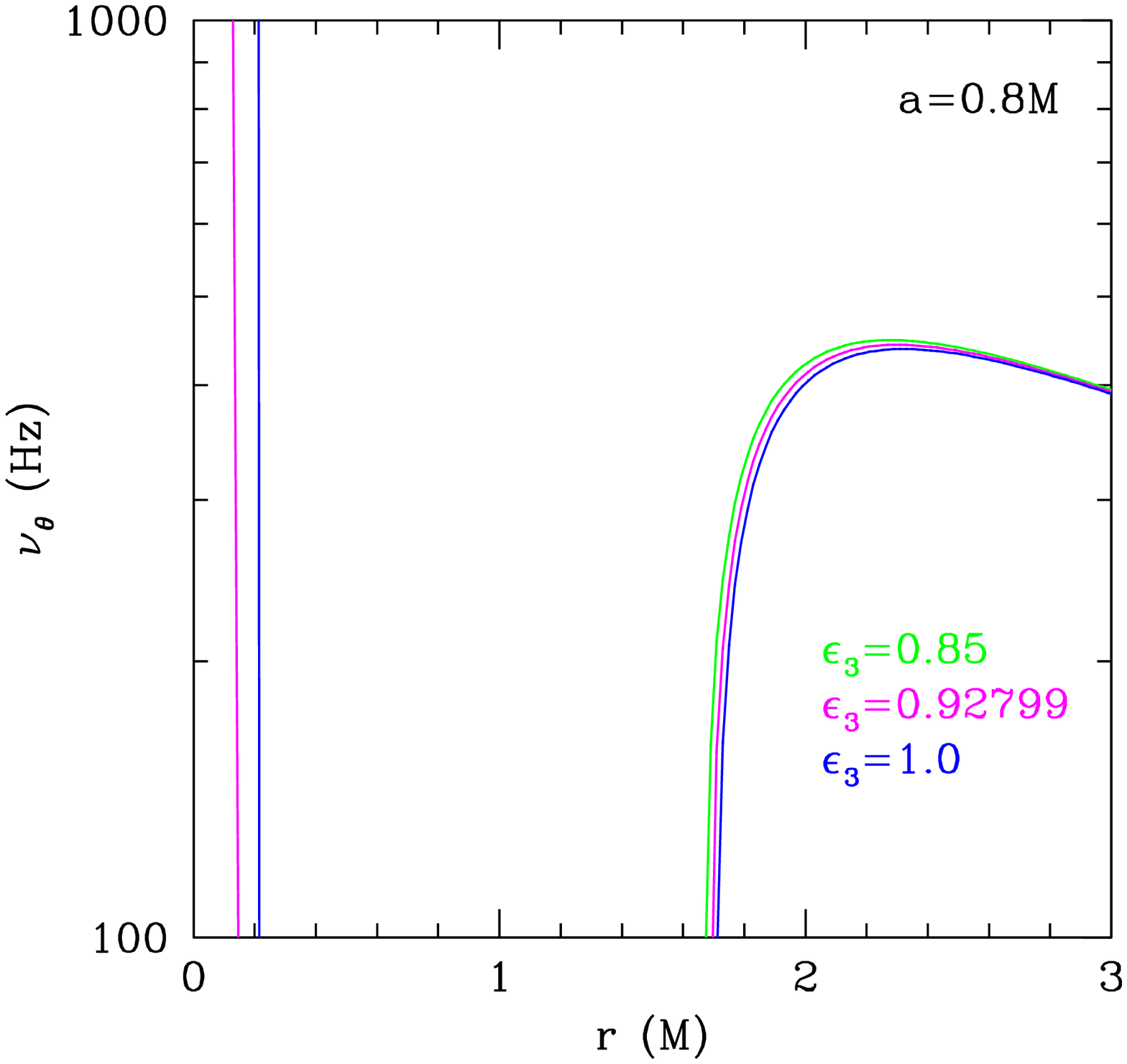,width=0.4\textwidth}
\end{center}
\caption{Radial (top) and vertical (bottom) epicyclic frequencies $\nu_r\equiv\Omega_r/2\pi$ and $\nu_\theta\equiv\Omega_\theta/2\pi$ for a $10M_\odot$ black hole for $a=0.8M$ and several values of the parameter $\epsilon_3$. If an ISCO exists, the radial epicyclic frequency vanishes at the ISCO radius, and circular equatorial orbits are radially unstable inside of this radius. If an ISCO does not exist, all circular orbits are radially stable but vertically unstable roughly in the interval $0.1M\lesssim r \lesssim1.7M$.}
\label{fig:freqs}
\end{figure}

In Fig.~\ref{fig:freqs}, I plot the radial and vertical epicyclic frequencies versus the radius for a spin value $a=0.8M$ and several values of the parameter $\epsilon_3$. For values of the parameter $\epsilon_3<\epsilon_3^{\rm bound}\approx0.92799$, circular orbits are radially unstable inside of the ISCO radius and are stable at and outside of the ISCO radius. For values of the parameter $\epsilon_3\geq\epsilon_3^{\rm bound}$, all circular orbits are radially stable but are vertically unstable over a small range of radii, where the vertical epicyclic frequency is imaginary.

\begin{figure}[ht]
\begin{center}
\psfig{figure=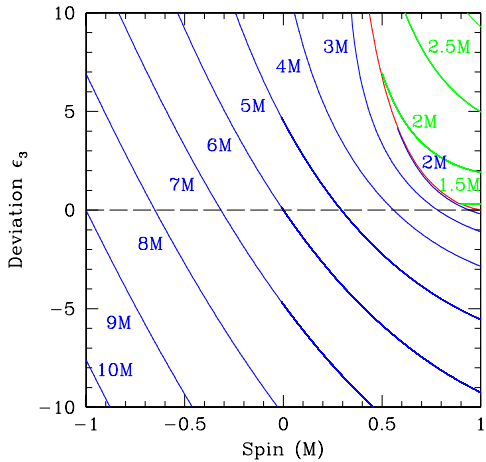,width=0.4\textwidth}
\end{center}
\caption{Contours of (blue curves) constant ISCO radius and (green curves) constant radius, at which the vertical epicyclic frequency vanishes, if no ISCO exists. The boundary between these regions (red curve) is determined by the function $\epsilon_3^{\rm bound}(a)$, $a>0$, given by Eq.~\eqref{eq:ep3bound}.}
\label{fig:ISCO}
\end{figure}

In Fig.~\ref{fig:ISCO}, I plot contours of constant ISCO radius as a function of the spin and the deviation parameter. An ISCO exists in the part of the parameter space where $a\leq0$ and where $\epsilon_3\lesssim\epsilon_3^{\rm bound}$, $a>0$. Contours of constant ISCO radius decrease for increasing values of the spin and decreasing values of the parameter $\epsilon_3$ \cite{JPmetric}. In the part of the parameter space, where $\epsilon_3\geq\epsilon_3^{\rm bound}$, $a>0$, I plot contours of the (outer) radii, at which the vertical epicyclic frequency vanishes causing circular equatorial orbits to become unstable in the vertical direction. These contours increase for increasing values of the spin and increasing values of the parameter $\epsilon_3$. The transition between radial and vertical instabilities of circular orbits across the boundary marked by the function $\epsilon_3^{\rm bound}(a)$, $a>0$, is generally not continuous; i.e., these instabilities occur at different radii just above and below this boundary. 

\begin{figure}[ht]
\begin{center}
\psfig{figure=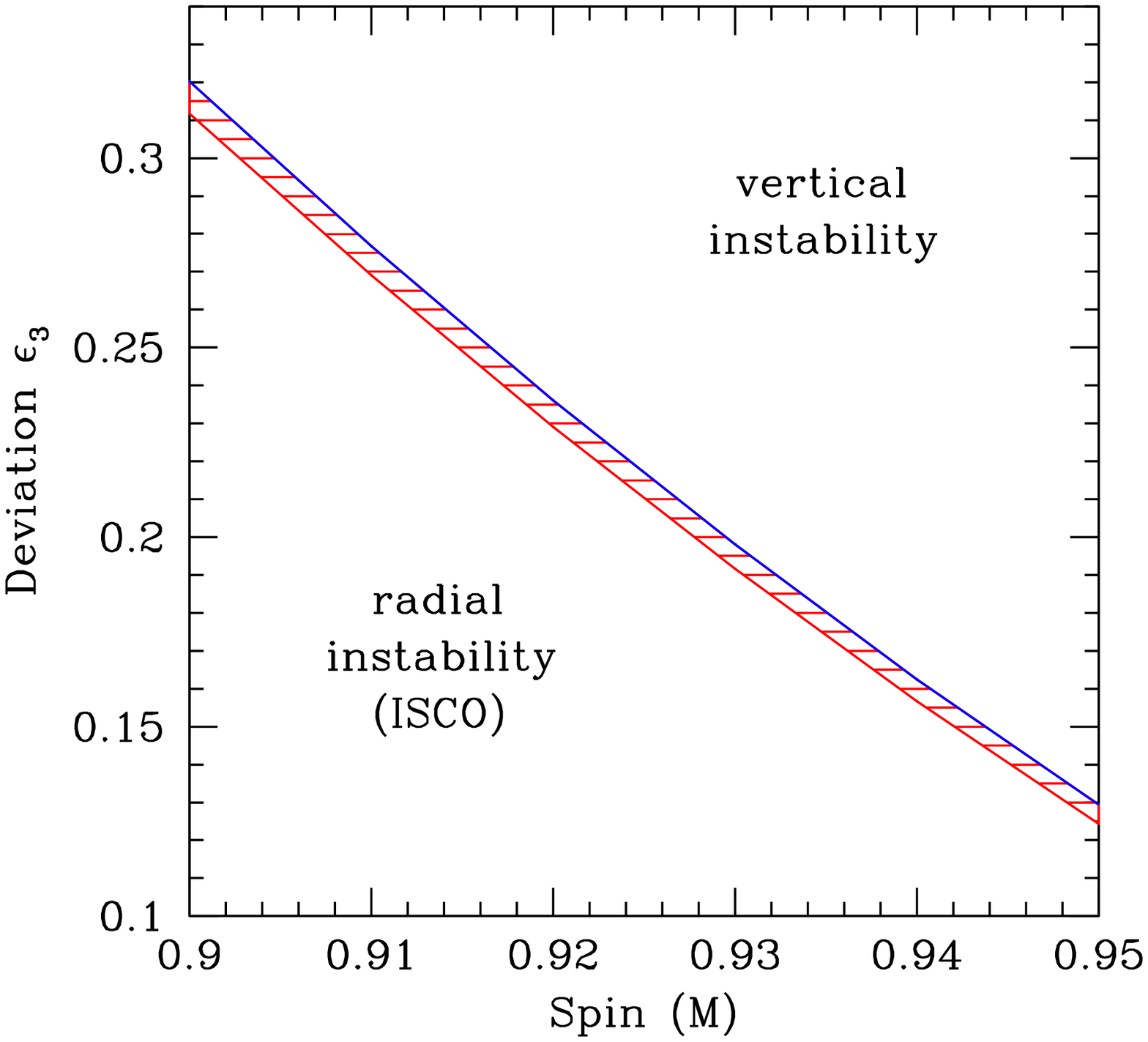,width=0.4\textwidth}
\psfig{figure=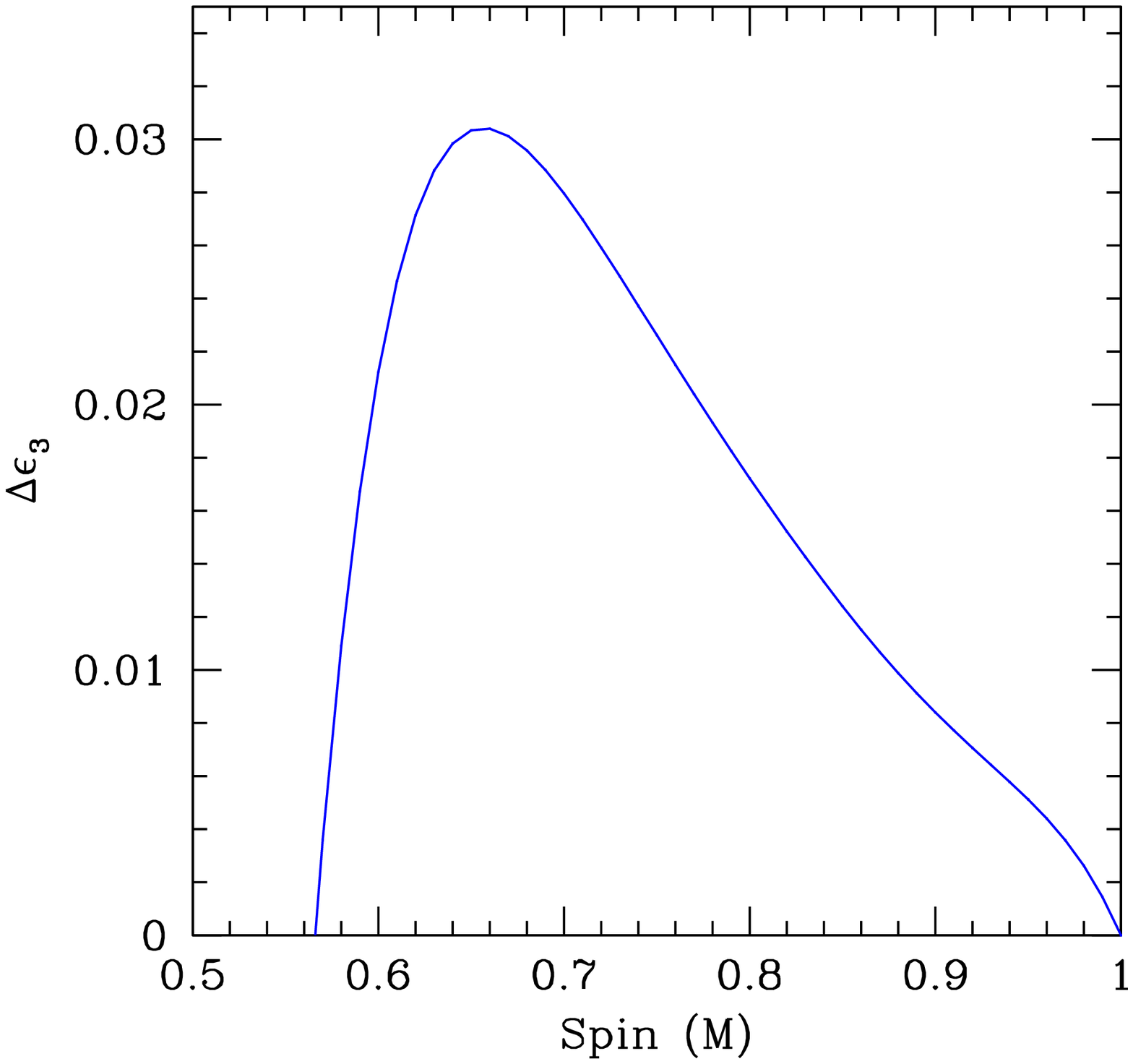,width=0.4\textwidth}
\end{center}
\caption{Top: Regions where an ISCO exists (labeled ``radial instability'') and where all circular equatorial orbits are radially stable but vertically unstable (labeled ``vertical instability''). The boundary between these regions (blue curve) is determined by the function $\epsilon_3^{\rm bound}(a)$, $a>0$, given by Eq.~\eqref{eq:ep3bound}. If an ISCO exists, circular equatorial motion is also vertically unstable, but this vertical instability only occurs inside of the ISCO except for values of the spin and the parameter $\epsilon_3$, which lie in a very narrow (red shaded) region along the boundary in the range $0.57M \lesssim a \leq M$. Bottom: Thickness $\Delta \epsilon_3$ of the red shaded region in the top panel.}
\label{fig:ISCOs}
\end{figure}

In the region of the parameter space, where an ISCO exists, circular equatorial motion is also unstable in the vertical direction. This instability, however, usually occurs inside of the ISCO and does not affect the structure of accretion disks. In a small and narrow region below the boundary $\epsilon_3^{\rm bound}(a)$, $a>0$, however, the radius where circular motion becomes vertically unstable lies slightly outside of the radius, where circular motion becomes radially unstable. This region lies in the spin range $0.57M\lesssim a < M$ and has an extent $\Delta \epsilon_3(a) \equiv \epsilon_3^{\rm bound}(a) - \epsilon_{3,{\rm ISCO}}(a) \lesssim 0.03$, where $\epsilon_{3,{\rm ISCO}}(a)$ is the value of the parameter $\epsilon_3$ at which the locations of the ISCO and the vertical instability coincide at a given value of the spin. I plot this region in Fig.~\ref{fig:ISCOs}. In Table~I, I summarize the properties of the inner edges of accretion disks and the different topologies of the Killing horizon.

\begin{table}[ht]
\begin{center}
\footnotesize
\begin{tabular}{ccccc}
\multicolumn{5}{c}{Killing Horizon and Inner Accretion Disk Edges}\\
\hline \hline

Spin & Deviation & Killing horizon & ISCO & Vertical   \\
     & parameter & topology        &      & instability \\
\hline
$a\leq0$ & $\epsilon_3<\epsilon_3^{\rm bound}$ & Spherical & Yes & Inside of ISCO \\
$a\leq0^{\rm a}$ & $\epsilon_3\geq\epsilon_3^{\rm bound}$ & Disjoint & Yes & Inside of ISCO \\
$a>0$ & $\epsilon_3<\epsilon_3^{\rm bound}$ & Spherical & Yes & Inside of ISCO$^{\rm b}$ \\
$a>0^{\rm a}$ & $\epsilon_3\geq\epsilon_3^{\rm bound}$ & Disjoint & No & Yes \\
\hline
\end{tabular}
\end{center}
\footnotetext{Kerr black hole if $a=\pm M$ with a spherical Killing horizon and the ISCO located at $r=M$ and $r=9M$, respectively.}
\footnotetext{Except for values of the spin and deviation parameter in a narrow region along the boundary $\epsilon_3^{\rm bound}$ (see Fig.~\ref{fig:ISCOs}).}
\caption{}
\end{table}

%%%%%%%%%%%%%%%%%%%%%%%%%%%%%%%%%%%%%%%%%%%%%%%%%%%%%

\section{Relativistically Broadened Iron Lines}
\label{sec:lines}

As a potential observational signature of the radiation emitted from accretion disks whose inner edges are determined by the vertical instability, I simulate profiles of relativistically broadened iron lines. I use the algorithm already developed in Ref.~\cite{PaperIV} with the proper implementation of the energy and axial angular momentum in the range where $\epsilon_3\geq\epsilon_3^{\rm bound}$, as listed in the Appendix.

As in Ref.~\cite{PaperIV}, I place an observer at a large distance $d$ from the central object and at an inclination $i$ from the rotation axis. I assume that the accretion disk is geometrically thin and optically thick and that it has a radial extent from the (outer) radius, where the vertical epicyclic frequency vanishes, to some outer radius $r_{\rm out}$. I assume that each particle in the disk moves on a circular Keplerian orbit, i.e., with the local Keplerian velocity $u=(u^t,0,0,u^\phi)$, where \cite{PaperIV}
\ba
u^t = - \frac{ g_{\phi\phi}E + g_{t\phi}L_z }{ g_{tt}g_{\phi\phi} - g_{t\phi}^2 }, \\
u^\phi = \frac{ g_{t\phi}E + g_{tt}L_z }{ g_{tt}g_{\phi\phi} - g_{t\phi}^2 }.
\label{velocity}
\ea
Further, I assume that the emission from the disk is monochromatic in the rest frame, with a typical energy $E_0\approx6.4~{\rm keV}$, and has an isotropic emissivity $\epsilon(r)\propto r^{-\alpha}$, where $\alpha$ is the emissivity index.

I calculate the photon paths to the distant observer via a fourth-order Runge-Kutta integration. These photons experience a redshift
\be
g \equiv \frac{E_{\rm im}}{E_{\rm d}} = \frac{ (g_{\mu\nu} k^\mu u^\nu)_{\rm im} }{ (g_{\mu\nu} k^\mu u^\nu)_{\rm d} },
\ee
where the subscripts ``im'' and ``d'' refer to the image plane and the accretion disk, respectively, and where $k^\mu$ is the 4-momentum of a given photon. The observed total specific flux is then given by the expression \cite{PaperIV}
\be
F_E = \frac{1}{d^2} \int d\alpha_0 \int d\beta_0 I(\alpha_0,\beta_0) \delta[ E_p - E_0 g(\alpha_0,\beta_0) ],
\ee
where $\alpha_0$ and $\beta_0$ are the (Cartesian) coordinates of the image plane and where $I(\alpha_0,\beta_0)$ and $E_p$ are the specific intensity and energy, respectively.

\begin{figure}[ht]
\begin{center}
\psfig{figure=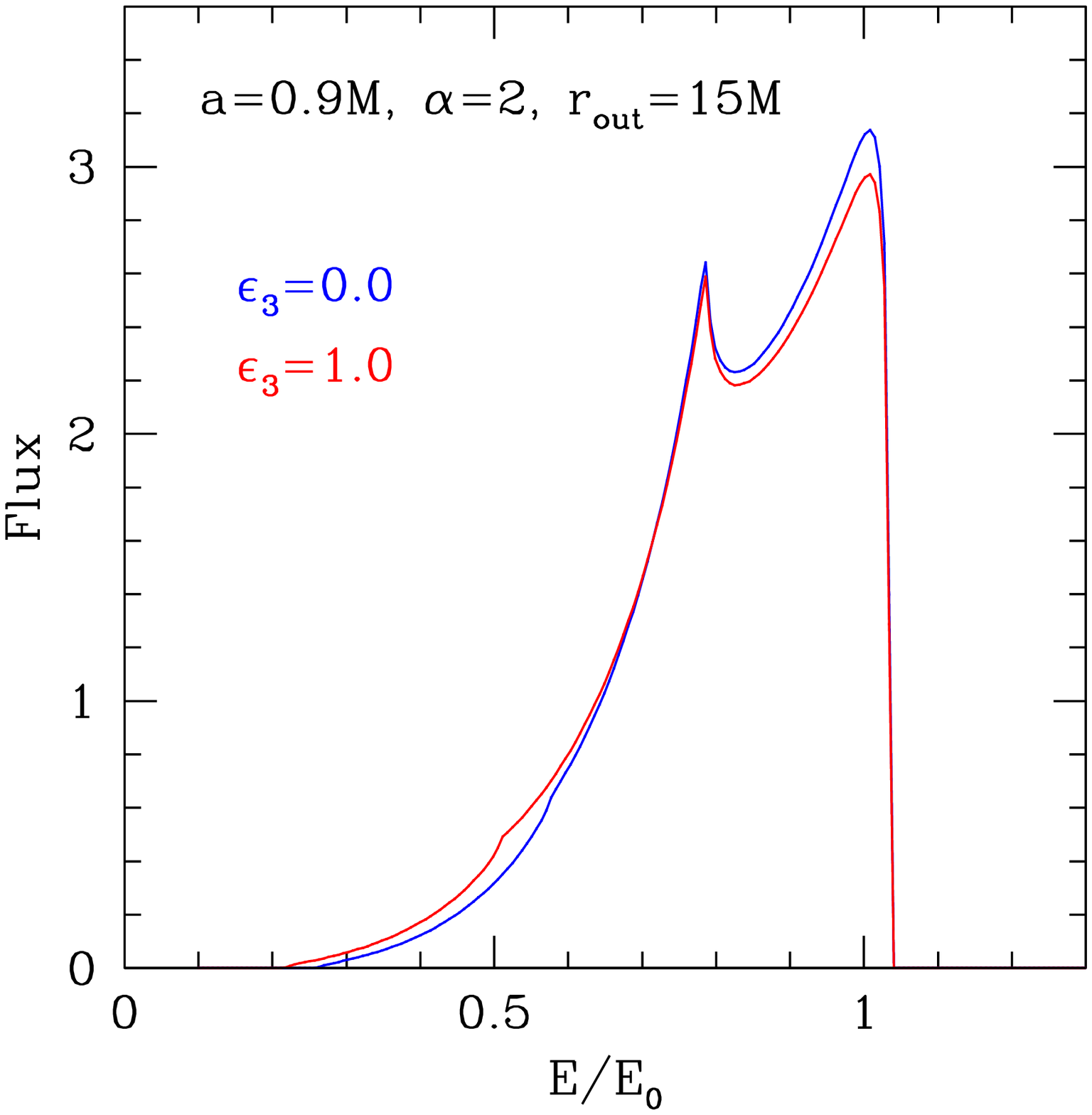,width=0.4\textwidth}
\psfig{figure=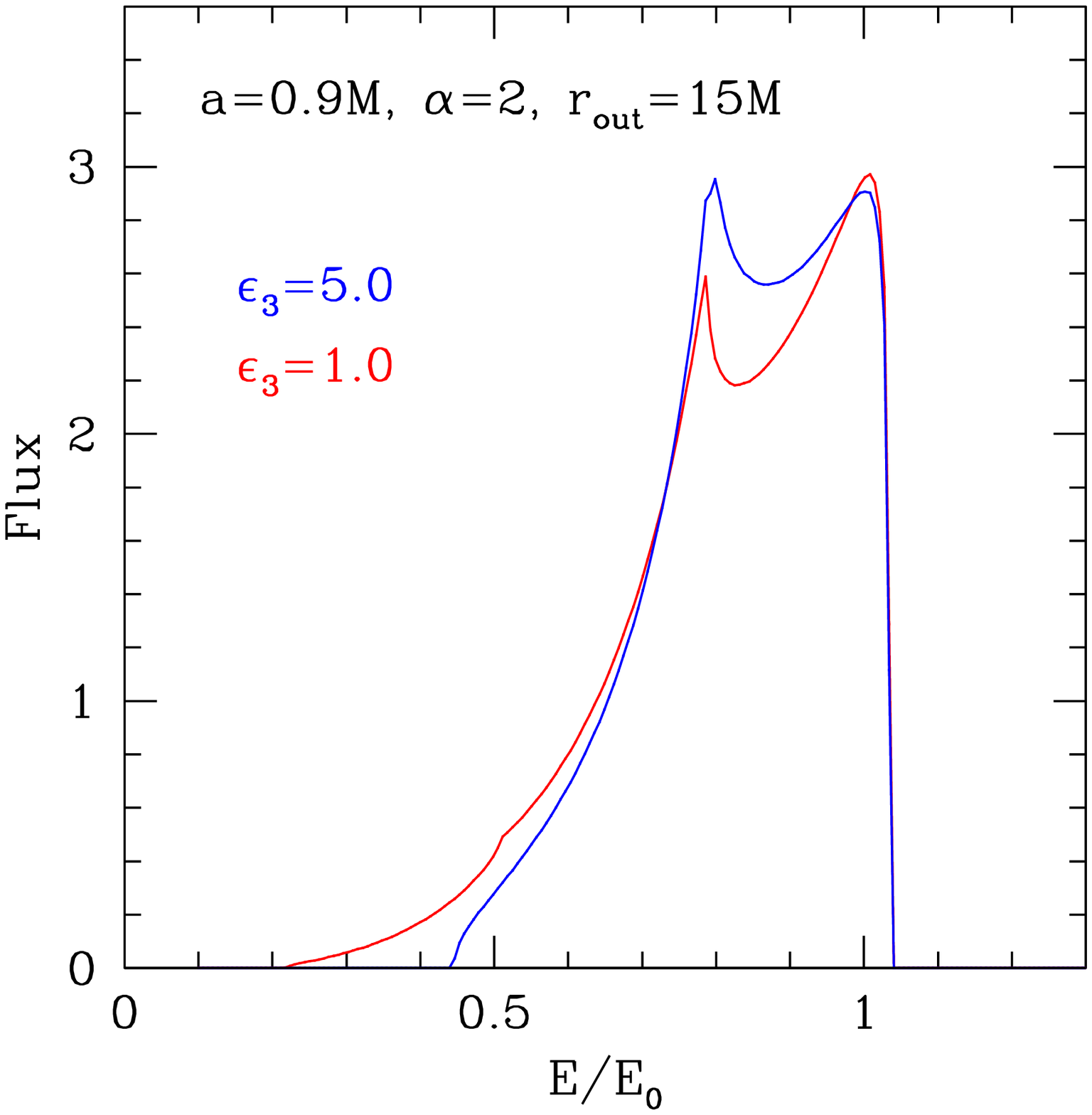,width=0.4\textwidth}
\end{center}
\caption{Profiles of relativistically broadened iron lines in units of the rest frame energy $E_0$ for values of the spin $a=0.9M$, emissivity index $\alpha=2$, outer disk radius $r_{\rm out}=15M$, disk inclination $i=30^\circ$, and several values of the deviation parameter. In the case of a Kerr black hole ($\epsilon_3=0$), the inner edge is located at the ISCO. For values of the parameter $\epsilon_3=1.0$ and $\epsilon_3=5.0$, the inner edge is located at the radius, at which the vertical epicyclic frequency vanishes. Changing the parameter $\epsilon_3$ primarily affects the flux ratio of the two peaks of a given profile as well as the low-energy spectrum corresponding to the location of the inner disk edge.}
\label{fig:ironline}
\end{figure}

In order to avoid any contact with the naked singularity located at the Killing horizon, I excise all radii $r<1.05\times r_{\rm hor}$, where $r_{\rm hor}$ is the maximal radius of the Killing horizon. This cutoff acts as an artificial event horizon, and any particle passing through this horizon is considered ``captured'' and does not contribute to the observed line flux. In principle, the flux emitted from the potential inner part of the accretion disk around the origin could also be included. In practice, however, if such an inner disk exists, the electromagnetic radiation emitted from this inner region likely accounts for only a small fraction of the total flux and is highly redshifted.

In Fig.~\ref{fig:ironline}, I plot iron line profiles for values of the spin $a=0.9M$, index $\alpha=2$, $r_{\rm out}=15M$, $i=30^\circ$, and several values of the deviation parameter. These lines have the characteristic double-horn profile with a ``blue'' peak at the energy $E\approx E_0$ and a ``red'' peak at a slightly lower energy, as well as a ``red tail'' at low energies which terminates at the energy with the highest gravitational redshift that corresponds to the emission from the inner accretion disk edge. The effect of changing the parameter $\epsilon_3$ is similar to its effect analyzed in Ref.~\cite{JPmetric}, which focused on values of the deviation parameter $\epsilon_3<\epsilon_3^{\rm bound}$. For increasing values of the parameter $\epsilon_3$, the flux of the blue peak diminishes, while the flux of the red peak increases and can even be larger than the flux of the blue peak, and the end of the red tail shifts according to the shift of the inner disk edge. In particular, changing the parameter $\epsilon_3$ affects the flux ratio of the two peaks, which is a potential observable. 

\begin{figure}[ht]
\begin{center}
\psfig{figure=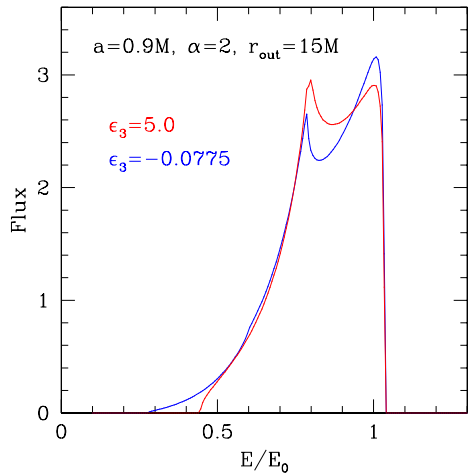,width=0.395\textwidth}
\end{center}
\caption{Profiles of relativistically broadened iron lines in units of the rest frame energy $E_0$ for values of the spin $a=0.9M$, emissivity index $\alpha=2$, outer disk radius $r_{\rm out}=15M$, disk inclination $i=30^\circ$, and two values of the deviation parameter. In both cases, the value of the deviation parameter is chosen so that the inner edge of the accretion disk as determined by either the ISCO ($\epsilon_3=-0.0775$) or the vertical instability ($\epsilon_3=5.0$) is located at the same radius ($r\approx2.423M$). The flux ratio of the peaks is much larger in the former case than in the latter case and the red tail extends to much lower energies.}
\label{fig:linesamerin}
\end{figure}

At a given value of the spin, the flux ratio of the two peaks can be significantly different for iron lines with different values of the deviation parameter, for which the ISCO and the radius of the vertical instability are in the same location. In Fig.~\ref{fig:linesamerin}, I plot iron lines with a spin parameter $a=0.9M$ and values of the deviation parameter $\epsilon_3=-0.0775$ and $\epsilon_3=5.0$, for which the inner accretion disk edge, as determined by either the ISCO or the vertical instability, respectively, is located at a radius $r\approx2.423M$. The flux ratio of the peaks is much larger when the inner edge is set by the ISCO ($\epsilon_3=-0.0775$) than when the inner edge is set by the radius of the vertical instability ($\epsilon_3=5.0$). In addition, in the former case the red tail is much more elongated, while it falls off abruptly in the latter case.

\begin{figure}[ht]
\begin{center}
\psfig{figure=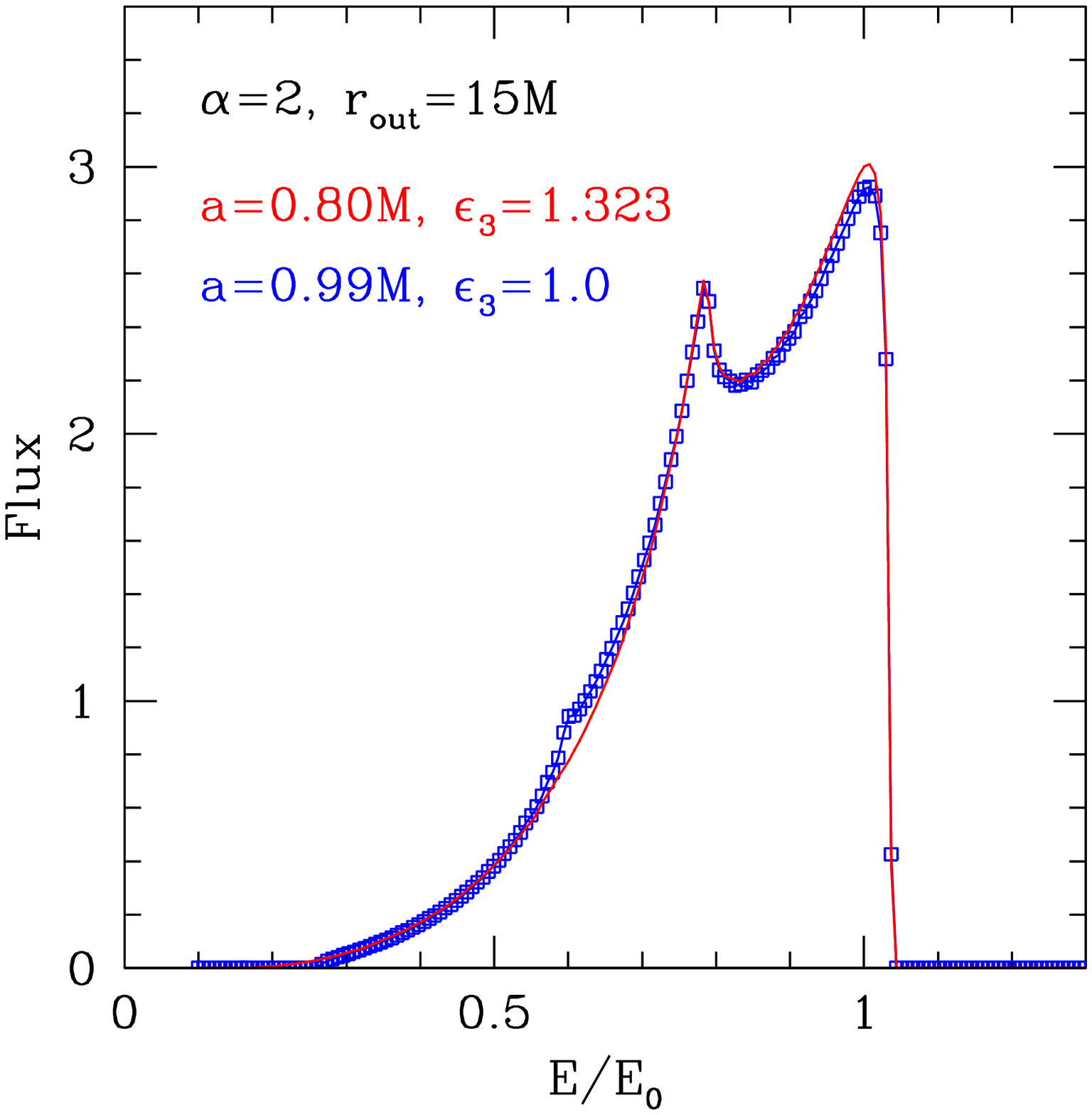,width=0.395\textwidth}
\psfig{figure=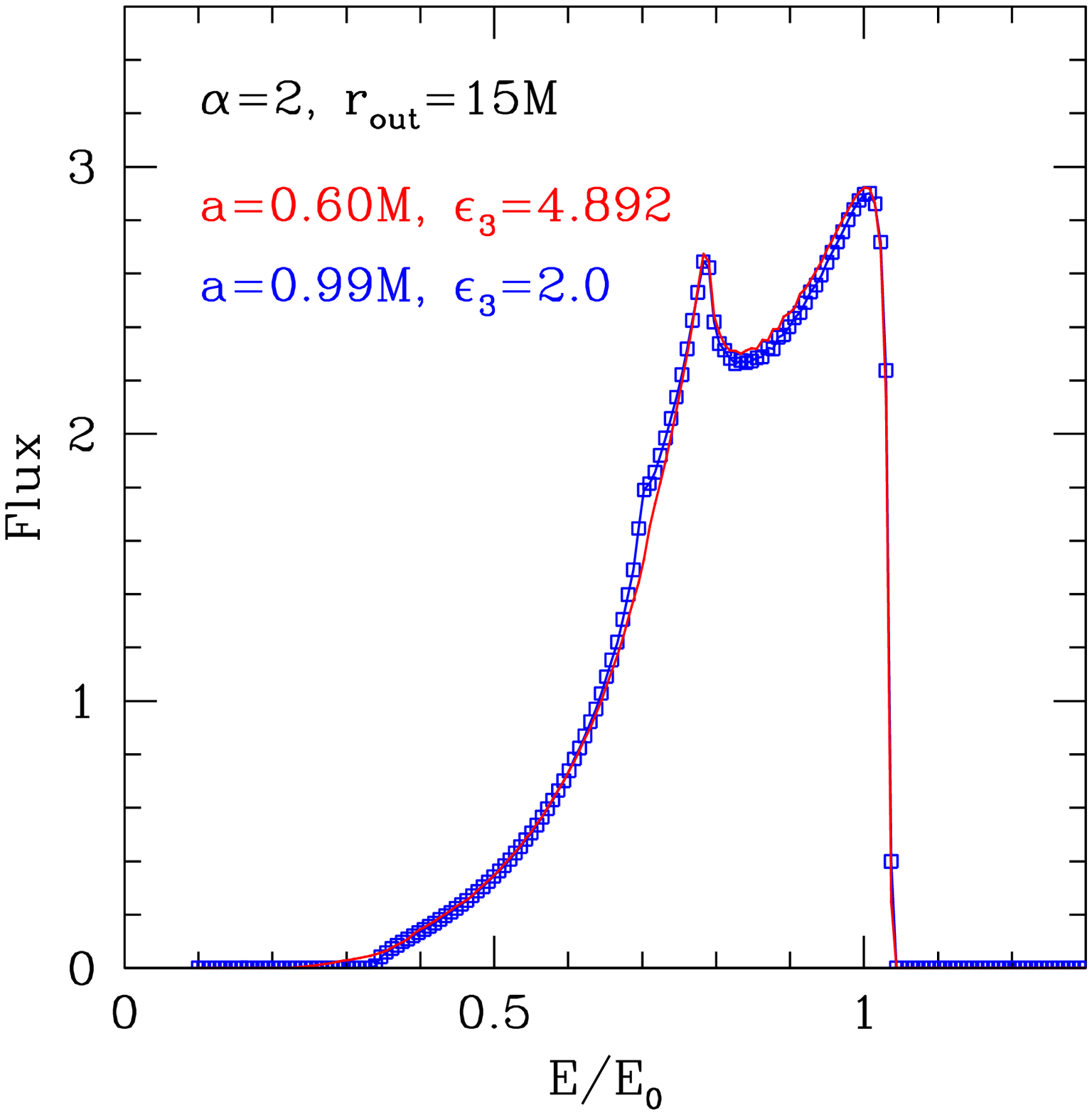,width=0.395\textwidth}
\end{center}
\caption{Iron line profiles in units of the rest frame energy $E_0$ with values of the spin and the deviation parameter chosen so that the inner disk edges coincide. The line profiles are very similar.}
\label{fig:sameISCO}
\end{figure}

As shown in Ref.~\cite{PaperIV}, profiles for sets of parameters with the same location of the ISCO are practically indistinguishable. In Fig.~\ref{fig:sameISCO}, I plot line profiles with two sets of parameters that correspond to the same radius of the vertical instability. In this case, the profiles are likewise very similar indicating a strong correlation between the locations of the inner edge and the line profiles.

%%%%%%%%%%%%%%%%%%%%%%%%%%%%%%%%%%%%
\section{Discussion}
\label{sec:discussion}

In this paper, I calculated the energy and axial angular momentum of a particle on a circular equatorial orbit as well as the radial and vertical epicyclic frequencies in the Kerr-like metric constructed by Johannsen and Psaltis \cite{JPmetric}. These functions depend explicitly on a set of functions $P_1$ to $P_6$ and their root structure, which I analyzed in detail.

I investigated the properties of the inner edges of thin accretion disks in this spacetime. While the ISCO is located at the radius, where the radial epicyclic frequency vanishes, an ISCO does not exist and circular equatorial orbits are radially stable at all radii in the range $\epsilon_3\geq \epsilon_3^{\rm bound}$, where the lower bound on the parameter $\epsilon_3$ is given by Eq.~\eqref{eq:ep3bound}. Instead, circular equatorial orbits become unstable in the vertical direction. I calculated the radii at which this instability occurs and showed that these locations depend significantly on the spin and the deviation parameter. The vertical instability exists even when an ISCO does exists, but I argued that the vertical instability plays no role in the structure of thin accretion disks in this case because it always lies inside of the ISCO unless the deviation parameter $\epsilon_3$ lies in a narrow region along the boundary $\epsilon_3^{\rm bound}(a)$, which I calculated explicitly. 

As a possible observable of accretion disks, which terminate at the radius where the vertical instability occurs, I simulated profiles of relativistically broadened iron lines and showed that these profiles depend significantly on the values of the spin and the deviation parameter. In particular, I showed that the flux ratio of the red and blue peaks can be significantly altered, which can serve as a probe of the deviation from the Kerr metric and which distinguishes iron lines with the same spin but unequal values of the deviation parameter for which the inner disk radius as determined by the ISCO and the vertical instability, respectively, is identical. As in the case of disks that terminate at the ISCO (see Ref.~\cite{PaperIV}), I also showed, however, that profiles for different choices of the spin and the deviation parameter that correspond to the same location of the inner edge as determined by the vertical instability are very similar.

\acknowledgments
This work was supported by a CITA National Fellowship at the University of Waterloo. Research at Perimeter Institute is supported by the Government of Canada through Industry Canada and by the Province of Ontario through the Ministry of Research and Innovation.

%%%%%%%%%%%%%%%%%%%%%%%%%%%%%%%%%%%%
%%%%%%%%%%%%%%%%%%%%%%%%%%%%%%%%%%%%
\appendix
\begin{widetext}
\section{ENERGY AND AXIAL ANGULAR MOMENTUM}

The functions $P_i$, $i=1,...,6$, that occur in the expressions of the energy and axial angular momentum of a particle on a circular equatorial orbit given by Eqs.~\eqref{eq:E} and \eqref{eq:Lz} have the form
\ba
P_1 = &&a^2 M r^4 \left(\epsilon_3 M^3 +r^3\right)^2 \bigg\{12\epsilon_3 a^2 M^3 \left(\epsilon_3 M^3+r^3\right)^2 \nonumber \\
&&-r^4 \left[2\epsilon_3 M^2 r^3 \left(3 r^2-8 M^2\right)+\epsilon_3^2 M^5 \left(40 M^2-48 M r+15 r^2\right)+4 r^6 (3 r-5 M)\right]\bigg\}, \\
P_2 = &&2 \bigg\{2 r^4\left(r^{20} \mp M P_4\right)+M r^{12} \bigg\{ 2 r^9 \left(-12 M^2+16 M r-7r^2\right) \nonumber \\
&&+\epsilon_3 M^2 (r-2 M)^2 \left[\epsilon_3^2 M^6 (5r-12 M)-6\epsilon_3 M^3 r^3 (5 M-2 r)-3 r^6 (8 M-3r)\right]\bigg\}\bigg\}, \\
P_3 = &&r^4 \left(12\epsilon_3 M^4 -5\epsilon_3 M^3 r +6 M r^3-2 r^4\right)^2-8 a^2 M \left(\epsilon_3 M^3 +r^3\right)^2 \left(5\epsilon_3 M^3 +2 r^3\right), \\
P_4 = &&
\begin{cases} -a B & \text{if }n_{41}<r<n_{42}
\\
a B &\text{else}
\end{cases}
, \\
P_5 = && M \left(r^3+\epsilon_3 M^3\right)^2 \bigg\{12\epsilon_3 a^6 M^3 \left(\epsilon_3 M^3 -2 r^3\right)^2 \left(\epsilon_3 M^3+r^3\right)^4+a^4 r^4 \left(\epsilon_3 M^3+r^3\right)^2 \nonumber \\
&&\left(-40\epsilon_3^4 M^{13}+40\epsilon_3^4 M^{12} r -15\epsilon_3^4 M^{11} r^2+128\epsilon_3^3 M^{10} r^3 -296\epsilon_3^3 M^9 r^4 +54\epsilon_3^3 M^8 r^5-924\epsilon_3^2 M^7 r^6 \right.\nonumber \\
&&\left.+276\epsilon_3^2 M^6 r^7 -36\epsilon_3^2 M^5 r^8 -880\epsilon_3 M^4 r^9 +304\epsilon_3 M^3 r^{10}-24\epsilon_3 M^2 r^{11}-112 M r^{12}+16 r^{13}\right) \nonumber \\
&&-2 a^2 r^8 \big[48 \epsilon_3^5 M^{17} -12\epsilon_3^5 M^{16} r -52\epsilon_3^5 M^{15} r^2+3\epsilon_3^4 M^{14} r^3 (5\epsilon_3+88)-720\epsilon_3^4 M^{13} r^4 +298\epsilon_3^4 M^{12} r^5 \nonumber \\
&& -3\epsilon_3^3 M^{11} r^6 (13 \epsilon_3+480)+516\epsilon_3^3 M^{10} r^7 +2\epsilon_3^3 M^9 r^8 -6\epsilon_3^2 M^8 r^9 (3\epsilon_3+508)+2292\epsilon_3^2 M^7 r^{10} \nonumber \\
&& -628 \epsilon_3^2 M^6 r^{11}+12\epsilon_3 M^5 r^{12} (5\epsilon_3-134)+1188\epsilon_3 M^4 r^{13} -296\epsilon_3 M^3 r^{14} +24 M^2 r^{15} (\epsilon_3-9)+120 M r^{16} \nonumber \\
&&-16r^{17}\big]-r^{14} \left(\epsilon_3 M^3 +6 M r^2-2 r^3\right)^2 \nonumber \\
&&\left(96\epsilon_3^2 M^7 -76\epsilon_3^2 M^6 r +15\epsilon_3^2 M^5 r^2+72\epsilon_3 M^4 r^3 -44\epsilon_3 M^3 r^4 +6\epsilon_3 M^2 r^5 +12 M r^6-4 r^7\right)\bigg\} \nonumber \\
&& -4 P_4 \left[ a^2 \left(\epsilon_3 M^3 -2 r^3\right)^2 \left(\epsilon_3 M^3+r^3\right)+6\epsilon_3 M^3 r^5 \left(\epsilon_3 M^3 +6 M r^2-2 r^3\right)\right], \\
P_6 = && -\epsilon_3^2 M^6 -6\epsilon_3 M^4 r^2 +\epsilon_3 M^3 r^3 -6 M r^5+2 r^6,
\ea
where
\be
B \equiv \sqrt{M \left(\epsilon_3 M^3 +r^3\right)^6 \left(9\epsilon_3^2 a^2 M^5+16\epsilon_3 M^3 r^4 -6\epsilon_3 M^2 r^5 +4 r^7\right) \left[a^2 \left(\epsilon_3 M^3 +r^3\right)+r^4 (r-2 M)\right]^2}
\ee
and where $n_{41}$ and $n_{42}$ denote the smaller and greater root of the function $P_4$, respectively, if real roots exist for $r>0$. Note the slight redefinition of the functions $P_4$ and $P_5$ with respect to the one in Ref.~\cite{JPmetric}.

Defining
\ba
E_+ &\equiv& \frac{1}{r^6}\sqrt{ \frac{P_1 + P_2}{P_3} }, \\
\label{eq:Eplus}
L_+ &\equiv& \frac{1}{r^4 P_6} \left[ \sqrt{\frac{P_5}{P_3} } - 6a M (r^3 + \epsilon_3 M^3) \sqrt{ \frac{P_1 + P_2}{P_3} } \right], \\
\label{eq:Lplus}
L_- &\equiv& -\frac{1}{r^4 P_6} \left[ \sqrt{\frac{P_5}{P_3} } + 6a M (r^3 + \epsilon_3 M^3) \sqrt{ \frac{P_1 + P_2}{P_3} } \right],
\label{eq:Lminus}
\ea
I can write the energy and axial angular momentum in the following manner:

\be
E =
\begin{cases}
E_+ & \text{if }r\geq n_{12} \\
& \text{if }r<n_{12} \begin{cases} E_+ & \text{if }\epsilon_3<\epsilon_3^{\rm bound} \\ -E_+ & \text{else} \end{cases}
\end{cases},
\ee

\ba
L_z &=&
\begin{cases}
  L_+ & \text{if }r>{\rm max}(n_3,n_5)
\\
      & \text{else}
    \begin{cases}
         & \text{if }r\geq n_5
       \begin{cases}
             & \text{if }n_3=n_5
           \begin{cases}
                & \text{if }r<n_3
              \begin{cases}
                 A & \text{if }r<n_{5a}

              \\
                   & \text{else}
                 \begin{cases}
                    L_+  & r\geq n_{12}
                 \\
                    -L_- & \text{else}
                 \end{cases}
              \end{cases}
           \\
                & \text{else}
              \begin{cases}
                 L_- & \text{if } {\rm sgn}^+={\rm sgn}^-
              \\
                 L_+ & \text{else}
              \end{cases}
           \end{cases}
       \\
         L_- & \text{else}
       \end{cases}
    \\
         & \text{else}

%\\
%      & \text{else}
    \begin{cases}
         & \text{if }r<n_3
       \begin{cases}
              & \text{if }r<n_{5a}
          \begin{cases}
             L_- & \text{if } {\rm sgn}^+={\rm sgn}^-
          \\
             L_+ & \text{else}
          \end{cases}
       \\
          L_+ & \text{else}
       \end{cases}
    \\
         & \text{else}
       \begin{cases}
          L_+ & \text{if } {\rm sgn}^+={\rm sgn}^-
       \\
              & \text{else}
          \begin{cases}
             L_+ & \text{if }n_3=n_5
          \\
             L_- & \text{else}
          \end{cases}
       \end{cases}
    \end{cases}
    \end{cases}
\end{cases},   \nn \\
{\rm where} \\
A &=&                  \begin{cases}
                       & {\rm sgn}^+={\rm sgn}^-
                    \begin{cases}
                           & \text{if }n\geq n_{12}
                       \begin{cases}
                         L_+ & \text{if }r\leq n_{5b}
                       \\
                         L_- & \text{else}
                       \end{cases}
                    \\
                      -L_- & \text{else}
                    \end{cases}
                 \\
                   L_+ & \text{else}
                 \end{cases}.
\ea

In these expressions, $n_{12}$ denotes the root of the function $P_1+P_2$, $n_3$ is the largest root of $P_3$, and $n_5$, $n_{5a}$, and $n_{5b}$ denote the roots of $P_5$, where $n_5\geq n_{5a}\geq n_{5b}$. Further, ${\rm max}(n_3,n_5)$ denotes the maximum of the roots $n_3$ and $n_5$, and ${\rm sgn}^\pm$ denote the signs of the function $P_5$ evaluated in the vicinity of the root $n_5$ at radii $r=n_5+\Delta r$ and $r=n_5-\Delta r$, respectively, where $\Delta r\ll M$. The latter definition is a simple check for the existence of either a single or a double root at $n_5$, which can easily be implemented in a numerical algorithm without having to compute the derivative of the function $P_5$. These expressions of the energy and axial angular momentum are valid for all values of the spin $|a|\leq M$ and the parameter $\epsilon_3$ if a circular equatorial orbit exists.

\end{widetext}

%%%%%%%%%%%%%%%%%%%%%%%%%%%%%%%%%%%%%%%%

\end{document}